\documentclass[twocolumn,showpacs,showkeys]{revtex4}
\usepackage{graphicx}
\usepackage{bm}
\usepackage{color}
\usepackage{amsmath}
\usepackage{natbib}

\begin{document}

\title{Spin electron acoustic soliton: Separate spin evolution of electrons with exchange interaction}

\author{Pavel A. Andreev}
\email{andreevpa@physics.msu.ru}
\affiliation{Faculty of physics, Lomonosov Moscow State University, Moscow, Russian Federation.}

 \date{\today}

\begin{abstract}
Separate spin evolution quantum hydrodynamics is generalized to include the Coulomb exchange interaction. The Coulomb exchange interaction is considered as the interaction between the spin-down electrons being in the quantum states occupied by one electron, giving main contribution in the equilibrium.
The generalized model is applied to study the non-linear spin-electron acoustic waves. Existence of the spin-electron acoustic soliton is demonstrated. Contributions of the concentration, spin polarization, and exchange interaction in the properties of the spin electron acoustic soliton are studied.
\end{abstract}


\maketitle



%


\section{\label{sec:level1} Introduction}

Spin evolution in quantum plasmas has been considered for a long
time. Fundamental equations of many-particle spin-1/2 quantum
hydrodynamics were derived in 2001 \cite{MaksimovTMP 2001},
\cite{MaksimovTMP 2001 b}. They found many applications at the
study of waves and instabilities in spin-1/2 quantum plasmas.
Among them we find the appearance of the spin-plasma waves with
frequencies near the cyclotron frequency \cite{Andreev VestnMSU
2007}, \cite{Brodin PRL 08 a}, \cite{Misra JPP 10}, \cite{Andreev
IJMP 12}. Later it was demonstrated that the quantum Bohm
potential existing in the magnetic moment evolution equation
shifts the frequency of the spin-plasma waves \cite{Trukhanova
PrETP 13}, \cite{Andreev arXiv 14 positrons}. This shift is
proportional to the square of the wave vector. The annihilation
interaction between electrons and positrons gives a shift of the
spin-plasma wave frequency on a constant proportional to the
magnetic moment density of the spin-1/2 quantum electron-positron
plasmas \cite{Andreev arXiv 14 positrons}. Generation of waves in
magnetized plasmas by neutron beams via the spin-spin and
spin-current interactions of the neutron spins with the spins and
electric currents of the electrons and ions of the plasmas was
considered in Ref. \cite{Andreev IJMP 12}. Spin parts of the
hydrodynamic vorticity and helicity of the spin-1/2 quantum
plasmas were derived in Refs. \cite{Mahajan PRL 11}, \cite{Mahajan
PL A 13}. Conservation of the full helicity at the charge-charge
and spin-spin interactions was demonstrated there. Spin parts of
the vorticity and helicity for spin-1/2 electron-positron plasmas
were obtained in Ref. \cite{Andreev arXiv 14 positrons}. These
phenomena appear along with the change of properties of well-known
plasma phenomena (see for instance Refs. \cite{Mushtaq PP 10},
\cite{Bychkov PP 10}, \cite{Chang Li PP 14}, \cite{Shukla RMP
11}).

Different methods of derivation \cite{Brodin NJP 07}, \cite{Koide
PRC 13} and generalization \cite{Andreev IJMP 12}, \cite{Andreev
RPJ 07}, \cite{Vagin 09}, \cite{Andreev IJMP B 15 spin current},
\cite{Andreev PRB 11} of the spin-1/2 quantum hydrodynamics were
presented in literature. Kinetic models for the spin-1/2 quantum
plasmas were presented as well \cite{Brodin PRL 08 a},
\cite{Andreev Physica A 15} (see also reviews \cite{Shukla RMP
11}, \cite{Uzdensky RPP 14}). All these models consider electrons
as a single fluid. It corresponds to the multi-fluid plasmas,
where each species is considered as a fluid. The Pauli equation
allows to find different form for the spin-1/2 quantum
hydrodynamics, where we have two fluids of electrons: the spin-up
electrons and the spin-down electrons \cite{Harabadze RPJ 04},
\cite{Brodin PRL 10 SPF}, \cite{Andreev PRE 15}. Early papers
\cite{Harabadze RPJ 04}, \cite{Brodin PRL 10 SPF} did not show
full picture of electron evolution and partially mistreat
coefficients in the spin-spin interaction force. The separate spin
evolution of the electrons, in accordance with the Pauli equation,
was obtained in Ref. \cite{Andreev PRE 15}.

It was demonstrated in Ref. \cite{Andreev PRE 15} that the Fermi pressures for the spin-up electrons and the spin-down electrons are different. It leads to existence of new phenomena. The spin-electron acoustic wave was found in Ref. \cite{Andreev PRE 15} at wave propagation parallel to the external magnetic field. Oblique propagation of the longitudinal waves at the separate evolution of the spin-up electrons and the spin-down electrons was considered in Ref. \cite{Andreev spin-up and spin-down 1406 Oblique}. Existence of two kinds of the spin-electron acoustic waves (SEAWs) was demonstrated in this regime. Properties of the SEAWs in two dimensional structures were studied in Ref. \cite{Andreev spin-up and spin-down 1408 2D}. The SEAWs in the two-dimensional plane-like electron gas and the electron gas on the cylindric surface were considered in Ref. \cite{Andreev spin-up and spin-down 1408 2D}. Kinetic model of the SEAWs was considered in Ref. \cite{Andreev spin-up and spin-down 1409 Kin}, where the Landau damping of the SEAWs was calculated. It was demonstrated that the Landau damping of the SEAWs is small. Therefore, the SEAWs are slowly damping waves.

Spins of electrons affect the plasma dynamics even if we do not consider spin evolution. It is enough to include distribution of electrons on different spin states to find a change in the equation of state. The distribution of electrons on spin states affects the Coulomb exchange interaction as well. The exchange interaction was considered at the first steps of the development of the many-particle spin-1/2 quantum hydrodynamics \cite{MaksimovTMP 2001 b}. The exchange interaction attracts a lot of attention in recent research (see for instance Refs. \cite{Zamanian PRE 13}-\cite{Andreev Trukhanova PoP 15}). This research are grounded on the long experience of the Coulomb exchange interaction study \cite{Nozieres PR 58}-\cite{Kohn RMP 99}, along with recent applications of the exchange interaction to different plasma phenomena (see for instance \cite{Jung PP 14}, \cite{Won Lee PP 13}).

In effort to study the influence of the Coulomb exchange
interaction on the properties of the spin-electron acoustic waves
we develop generalization of the separate spin evolution quantum
hydrodynamics (SSE-QHDs) \cite{Andreev PRE 15} containing
contribution of the exchange interaction. As in Ref. \cite{Andreev
PRE 15} we focus our attention on degenerate electron gas.

In this paper we consider the Coulomb exchange interaction, which arises as the interaction between the spin-down electrons being in the quantum states occupied by one electron. It is considered in Ref. \cite{Andreev AoP 14} in terms of single fluid model of electrons. One of the features of this paper is the application of the exchange interaction obtained in Ref. \cite{Andreev AoP 14} to two-fluid model of electrons \cite{Andreev PRE 15}.

We apply the developed model to the non-linear SEAWs, particularly
to the soliton formation.

Non-linear waves related to the ion-acoustic waves are still under
consideration \cite{Hanif PP 14}, while this paper is dedicated to
non-linear waves related to the recently found spin-electron
acoustic waves \cite{Andreev PRE 15}.

This paper is organized as follows. In Sec. II we present the QHD
model with separated spin-up electrons and spin-down electrons
containing the Coulomb exchange interaction. In Sec. III we
describe method of derivation of the spin-electron acoustic
soliton from the developed in Sec. II model. In Sec. IV we present
analysis of properties of the  spin-electron acoustic soliton in
quantum plasmas with no account of the exchange interaction. In
Sec. V we describe the spin-electron acoustic soliton with the
account of the exchange interaction. In Sec. VI brief summary of
obtained results is presented.

\section{\label{sec:level1} Model}

The Pauli equation is a set of two equations describing evolution
of two wave functions, one is for spin-up state of electron and
another one is for spin-down state of electron. Therefore, the
evolution of system of electrons can be described in terms of
two-fluid model of electrons with different spin projection. This
is called the separate spin evolution quantum hydrodynamics
\cite{Andreev PRE 15}, \cite{Andreev 1410 Comment}. Corresponding kinetic model is obtained as
well \cite{Andreev spin-up and spin-down 1409 Kin}. However these
models are derived in the self-consistent field approximation. In
this paper we make the next step in the development of the
SSE-QHD. We include the Coulomb exchange interaction in the
SSE-QHD.

The time evolution of the concentrations of the spin-up electrons and the spin-down electrons obeys the continuity equations with nonzero right-hand side
\begin{equation}\label{SEAS EX cont eq electrons spin UP GEN}
\partial_{t}n_{\uparrow}+\nabla(n_{\uparrow}\textbf{v}_{\uparrow})=\frac{\mu_{e}}{\hbar}(S_{x}B_{y}-S_{y}B_{x}), \end{equation}
and
\begin{equation}\label{SEAS EX cont eq electrons spin DOWN GEN}
\partial_{t}n_{\downarrow}+\nabla(n_{\downarrow}\textbf{v}_{\downarrow})=-\frac{\mu_{e}}{\hbar}(S_{x}B_{y}-S_{y}B_{x}), \end{equation}
where $n_{\uparrow}$ and $n_{\downarrow}$ ($\textbf{v}_{\uparrow}$
and $\textbf{v}_{\downarrow}$) are the particle concentrations
(the velocity fields) of the spin-up electrons and the spin-down
electrons, $\mu_{e}=-g\frac{e\hbar}{2mc}$ is the magnetic moment of electron, and $g=1+\alpha/(2\pi)=1.00116$, where
$\alpha=1/137$ is the fine structure constant, which gets into
account the anomalous magnetic moment of electron, $e$ ($m_{e}$)
is the electron charge (mass), $\hbar$ is the Planck constant, $c$
is the speed of light, $\textbf{B}=\{ B_{x}, B_{y}, B_{z}\}$ is
the magnetic field. The particle concentrations appear as the
quantum mechanical average of the corresponding wave functions,
which are the elements of the Pauli spinor wave function,
$n_{s}=\langle\psi_{s}^{*}\psi_{s}\rangle$, with $s=\uparrow$ or
$\downarrow$. These concentrations are related to the spin-up
electrons and the spin-down electrons separately. It appears in
the accordance with the spinor structure of the Pauli equation,
which governs the evolution of the spinor wave function
$\psi=(\begin{array}{c}
                                                                                                                  \psi_{\uparrow} \\
                                                                                                                  \psi_{\downarrow}
                                                                                                                \end{array}
)$. Considering the evolution of the upper and lower elements
separately we find the separate description of the spin-up
electrons and the spin-down electrons. The velocity fields
$\textbf{v}_{\uparrow}$ and $\textbf{v}_{\downarrow}$ appear at
the averaging of the corresponding operators with
$\psi_{\uparrow}$ and $\psi_{\downarrow}$:
$\textbf{v}_{s}=(1/n_{s})\langle(\psi_{s}^{*}\textbf{p}\psi_{s}+c.c.)/2m_{e}\rangle$,
where c.c. stands for the complex conjugation. Equations
(\ref{SEAS EX cont eq electrons spin UP GEN}) and (\ref{SEAS EX
cont eq electrons spin DOWN GEN}) contain projections of the spin
density $S_{x}$ and $S_{y}$. Each projection of the spin density
is defined as a mixture of the spin-up and spin-down wave
functions:
$S_{x}=\psi^{*}\sigma_{x}\psi=\psi_{\downarrow}^{*}\psi_{\uparrow}+\psi_{\uparrow}^{*}\psi_{\downarrow}$,
and
$S_{y}=\psi^{*}\sigma_{y}\psi=\imath(\psi_{\downarrow}^{*}\psi_{\uparrow}-\psi_{\uparrow}^{*}\psi_{\downarrow})$.
Therefore, these quantities are not related to different species of
electrons having different spin direction. $S_{x}$ and $S_{y}$
describe simultaneous evolution of both species of electrons.
Equations of evolution of $S_{x}$ and $S_{y}$ were derived in Ref.
\cite{Andreev PRE 15} as a part of the set of SSE-QHD equations.
We do not study the spin evolution in this paper, so we do not
describe equations for $S_{x}$ and $S_{y}$, which can be found in
Refs. \cite{Andreev PRE 15} and \cite{Andreev spin-up and
spin-down 1406 Oblique}.

From the continuity equations (\ref{SEAS EX cont eq electrons spin UP GEN}) and (\ref{SEAS EX cont eq electrons spin DOWN GEN}) we see that the numbers of electrons in each subspecies can change due to the spin-spin interaction and the interaction of spins with the external magnetic field. However, the full number of electrons $n_{e}=n_{\downarrow}+n_{\uparrow}$ conserves in this model.

In this model we have two Euler equations. We use the subindex $s=\uparrow$ or $\downarrow$ to present them as one equation
$$mn_{s}(\partial_{t}+\textbf{v}_{s}\nabla)\textbf{v}_{s}+\nabla p_{s}-\frac{\hbar^{2}}{4m}n_{s}\nabla\Biggl(\frac{\triangle n_{s}}{n_{s}}-\frac{(\nabla n_{s})^{2}}{2n_{s}^{2}}\Biggr)$$
$$=q_{e}n_{s}\biggl(\textbf{E}+\frac{1}{c}[\textbf{v}_{s},\textbf{B}]\biggr)+\textbf{F}_{Ex,s}$$
$$\pm\mu_{e}n_{s}\nabla B_{z} +\frac{\mu_{e}}{2}(S_{x}\nabla B_{x}+S_{y}\nabla B_{y})$$
\begin{equation}\label{SEAS EX Euler eq electrons spin UP GEN} \pm\frac{m\mu_{e}}{\hbar}(\textbf{J}_{(M)x}B_{y}-\textbf{J}_{(M)y}B_{x})
\mp m\textbf{v}_{\uparrow}\frac{\mu_{e}}{\hbar}(S_{x}B_{y}-S_{y}B_{x}),\end{equation}
where in coefficients $\pm$ and $\mp$ we have the upper sign for the spin-up electrons and the lower one for the spin-down electrons. In formula (\ref{SEAS EX Euler eq electrons spin UP GEN}) we use $q_{e}=-e$ for the electron charge, $p_{s}$ for the pressure of the spin-up and spin-down electrons. We also apply  $\textbf{J}_{(M)x}$ and $\textbf{J}_{(M)y}$ for the elements of the spin current tensor $J^{\alpha\beta}$. Vectors $\textbf{J}_{(M)x}$ and $\textbf{J}_{(M)y}$ have the following explicit forms
\begin{equation}\label{SEAS EX Spin current x} \textbf{J}_{(M)x}=\frac{1}{2}(\textbf{v}_{\uparrow}+\textbf{v}_{\downarrow})S_{x}-\frac{\hbar}{4m} \biggl(\frac{\nabla n_{\uparrow}}{n_{\uparrow}}-\frac{\nabla n_{\downarrow}}{n_{\downarrow}}\biggr)S_{y}, \end{equation}
and
\begin{equation}\label{SEAS EX Spin current y} \textbf{J}_{(M)y}= \frac{1}{2}(\textbf{v}_{\uparrow}+\textbf{v}_{\downarrow})S_{y} +\frac{\hbar}{4m}\biggl(\frac{\nabla n_{\uparrow}}{n_{\uparrow}}-\frac{\nabla n_{\downarrow}}{n_{\downarrow}}\biggr)S_{x}. \end{equation}

The Euler equations (\ref{SEAS EX Euler eq electrons spin UP GEN}) describe the momentum evolution complicated by the unconservation of the numbers of the spin-up and spin-down electrons.

We describe now the physical meaning of different terms in the Euler equations (\ref{SEAS EX Euler eq electrons spin UP GEN}). The first term is the continual derivative of the velocity field $\textbf{v}_{s}$. The second term is the gradient of pressure. The explicit form of the pressure we present and discuss below. The third term is proportional to the square of the Planck constant. It is a combination of the spatial derivatives of the particle concentration $n_{s}$ up to the third derivative $\nabla\triangle n_{s}$. This term is called the quantum Bohm potential. It is related to the wave nature of the electron.

On the right-hand side of the Euler equations (\ref{SEAS EX Euler
eq electrons spin UP GEN}) we present the force fields of
different nature. The first term presents the Lorentz force
describing  the interaction of charges with the electromagnetic
fields. The electric field contains two parts
$E=-\nabla\phi-\partial_{t}\textbf{A}/c$, where $\phi$ and
$\textbf{A}$ are the scalar and vector potentials of the
electromagnetic field. The first of them is the potential part
giving contribution in the longitudinal waves, which we consider
in this paper, and the second one is the vortical part. The second
part of the Lorentz force describes the interaction of the moving
charges with the magnetic field. The second term on the right-hand
side of the Euler equations (\ref{SEAS EX Euler eq electrons spin
UP GEN}) is the exchange part of the Coulomb interaction.   The
contribution of the Coulomb exchange interaction in the separate
spin evolution QHD is in the center of attention of this paper. We
discuss its explicit form below.

The third and fourth terms are the parts of the spin-spin
interaction force. The magnetic moments related to the spin of
electrons create the magnetic field. This magnetic field acts on
the magnetic moments of other electrons and leads to existence of
this force in the Euler equation. In terms of the SSE-QHD this
force field splits on two terms. The first (the second) of them
describes the interaction of the z projection (the x- and y
projections) of the magnetic moments with the nonuniform z
projection (the x- and y projections) of the magnetic field. The
last two terms on the right-hand side of the Euler equations
(\ref{SEAS EX Euler eq electrons spin UP GEN}) are related to the
unconservation of the numbers of the spin-up and spin-down
electrons. The first of them appears at the derivation of the
Euler equation for the momentum density $n_{s}\textbf{v}_{s}$. The
second of them arises at the application of the continuity
equation during the extraction of $\partial_{t}\textbf{v}_{s}$
from $\partial_{t}(n_{s}\textbf{v}_{s})$. Hence, it is
proportional to the right-hand side of corresponding continuity
equation.

\begin{figure}
\includegraphics[width=8cm,angle=0]{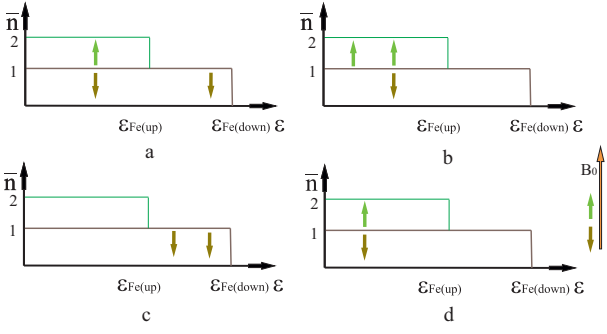}
\caption{\label{SEAS EX_Fermi_step} (Color online) The figure shows the different combinations of pair of the spin-up and spin-down electrons in the degenerate electron gas giving different contributions in the collective effect of the Coulomb exchange interaction. In this figure we apply the Fermi energies of the spin-up and spin-down electrons: $\varepsilon_{Fe(up)}=(6\pi^{2}n_{0\uparrow})^{2/3}\hbar^{2}/2m$, $\varepsilon_{Fe(down)}=(6\pi^{2}n_{0\downarrow})^{2/3}\hbar^{2}/2m$.}
\end{figure}

Study of the Coulomb exchange interaction in the electron gas has
a long history. Recently, it was shown that the exchange
interaction force field strongly depends on the spin polarization
of the electron gas \cite{Andreev AoP 14}. Distribution of the
partially spin polarized electrons is depicted in Fig. \ref{SEAS
EX_Fermi_step}. This distribution splits the electrons on
different groups, which demonstrate different exchange
interactions. Fig. (\ref{SEAS EX_Fermi_step} a) shows two pairs of
electrons. We see a spin-down electron being in a state with
energy $E\in(\varepsilon_{Fe(up)},\varepsilon_{Fe(down)}]$
interacting with two electrons having different spin direction and
being in the same quantum state with energy
$E'\in[0,\varepsilon_{Fe(up)}]$, where
$\varepsilon_{Fe(up)}=(6\pi^{2}n_{0\uparrow})^{2/3}\hbar^{2}/2m$,
$\varepsilon_{Fe(down)}=(6\pi^{2}n_{0\downarrow})^{2/3}\hbar^{2}/2m$.
The strengths of these interactions are the same, but they have
opposite signs. Hence, it gives zero contribution in the force
field.

In Fig. (\ref{SEAS EX_Fermi_step} b) we have a similar situation, but a chosen spin-up electron is in a quantum state with energy $E\in[0,\varepsilon_{Fe(up)}]$. It gives zeroth contribution in the force field either. If we consider a spin-down electron in a quantum state with energy $E\in[0,\varepsilon_{Fe(up)}]$ we find the same result.

Fig. (\ref{SEAS EX_Fermi_step} c) shows the regime giving non zero contribution in the force field. The regime presented in Fig. (\ref{SEAS EX_Fermi_step} c) was considered in Ref. \cite{Andreev AoP 14}. Here, we have interaction of two spin-down electrons being in quantum states with energies $E\in(\varepsilon_{Fe(up)},\varepsilon_{Fe(down)}]$. In this regime the spatial part of the wave function is antisymmetric relatively permutation of these two particles. We have same sign of interaction for all such pairs of electrons. It is important to underline that this regime involves the interaction of spin-down electrons only. Hence we should substitute this force field, found in Ref. \cite{Andreev AoP 14}, in the Euler equation for the spin-down electrons. This regime gives no contribution in the Euler equation for the spin-up electrons.

The regime of the electron interaction depicted in Fig. (\ref{SEAS
EX_Fermi_step} d) describes the
electrons having opposite spins and located in the same quantum
state. In this case the spatial part of the wave function is
symmetric relatively to the permutation of two particles even without additional symmetrization $\psi_{\textbf{p}_{i}\textbf{p}_{i}}(\textbf{r}_{1},\textbf{r}_{2})=\psi_{\textbf{p}_{i}}(\textbf{r}_{1})\psi_{\textbf{p}_{i}}(\textbf{r}_{2}) =\psi_{\textbf{p}_{i}\textbf{p}_{i}}(\textbf{r}_{1},\textbf{r}_{2})$. Therefore, there is no exchange interaction in this regime.

The force of the exchange interaction of the spin-down electrons being in quantum states occupied by one electron was found in Ref. \cite{Andreev AoP 14}. The result was presented in terms of the concentration of all electrons, while it involves the interaction of spin-down electrons only. To substitute this force in the Euler equation for the spin-down electrons it is necessary to rewrite the force field in terms of the spin-down electron concentration. In the equilibrium state we have the following relations between the spin polarization $\eta$, the spin-up electron concentration $n_{\uparrow}$, the spin-down electron concentration $n_{\downarrow}$, and the full concentration of electrons $n_{e}$:
$n_{\uparrow}=(1-\eta)n_{e}/2$,
$n_{\downarrow}=(1+\eta)n_{e}/2$.
Applying these formulae we can make the required representation of the exchange interaction force:
\begin{equation}\label{SEAS EX F EX parallel}\textbf{F}_{Ex,\downarrow\downarrow}=\zeta_{3D} q_{e}^{2}\sqrt[3]{\frac{3}{\pi}}\sqrt[3]{n_{e}}\nabla n_{e}= \chi q_{e}^{2}\sqrt[3]{n_{\downarrow}}\nabla n_{\downarrow},\end{equation}
where
\begin{equation}\label{SEAS EX zeta def} \zeta_{3D}=(1+\eta)^{4/3}-(1-\eta)^{4/3},\end{equation}
and
\begin{equation}\label{SEAS EX} \chi=\zeta_{3D} \sqrt[3]{\frac{3}{\pi}} \frac{2^{\frac{4}{3}}}{(1+\eta)^{4/3}} =2^{\frac{4}{3}}\sqrt[3]{\frac{3}{\pi}}\biggl(1-\frac{(1-\eta)^{4/3}}{(1+\eta)^{4/3}}\biggr).\end{equation}

The electromagnetic field presented in the hydrodynamic equations
satisfy the Maxwell equations:
\begin{equation}\label{SEAS EX div E full} \nabla \textbf{E}=4\pi(en_{i}-en_{e\uparrow}-en_{e\downarrow}),\end{equation}
\begin{equation}\label{SEAS EX div B full} \nabla \textbf{B}=0, \end{equation}
\begin{equation}\label{SEAS EX rot E full} \nabla\times \textbf{E}=-\frac{1}{c}\partial_{t}\textbf{B},\end{equation}
and
$$\nabla\times \textbf{B}=\frac{1}{c}\partial_{t}\textbf{E}$$
\begin{equation}\label{SEAS EX rot B full}
+\frac{4\pi}{c}\sum_{a=e,i}(q_{a}n_{a\uparrow}\textbf{v}_{a\uparrow}+q_{a}n_{a\downarrow}\textbf{v}_{a\downarrow})+4\pi\sum_{a=e,i}\nabla\times \textbf{M}_{a},\end{equation}
where $\textbf{M}_{a}=\{\mu_{a}S_{ax}, \mu_{a}S_{ay}, \mu_{a}(n_{a\uparrow}-n_{a\downarrow})\}$ is the magnetization of electrons in terms of hydrodynamic variables.

We consider the wave propagation parallel to the external magnetic field. Since we consider the longitudinal waves $\textbf{k}\parallel \delta \textbf{E}$ we find from equation (\ref{SEAS EX rot E full}) that the perturbation of the magnetic field is equal to zero $\delta \textbf{B}=0$. Hence the set of SSE-QHD equations (\ref{SEAS EX cont eq electrons spin UP GEN})-(\ref{SEAS EX Spin current y}) simplifies to the following set of equations.

We consider interval of the equilibrium concentrations from $n_{0}=10^{21}$ cm$^{-3}$ to $n_{0}=10^{27}$ cm$^{-3}$.
We drop the contribution of the quantum Bohm potential, which reveals itself at larger concentrations.

In this regime we have conservation of the electron number for both subspecies
\begin{equation}\label{SEAS EX cont eq electrons spin UP}
\partial_{t}n_{\uparrow}+\nabla(n_{\uparrow}\textbf{v}_{\uparrow})=0, \end{equation}
and
\begin{equation}\label{SEAS EX cont eq electrons spin DOWN}
\partial_{t}n_{\downarrow}+\nabla(n_{\downarrow}\textbf{v}_{\downarrow})=0. \end{equation}
The conservation follows from the zeroth value on the right-hand sides of the continuity equations (\ref{SEAS EX cont eq electrons spin UP}) and (\ref{SEAS EX cont eq electrons spin DOWN}).

Simplified Euler equations appear as follows
$$mn_{\uparrow}(\partial_{t}+\textbf{v}_{\uparrow}\nabla)\textbf{v}_{\uparrow}+\nabla p_{\uparrow}$$
\begin{equation}\label{SEAS EX Euler eq electrons spin UP} =q_{e}n_{\uparrow}\textbf{E}+\frac{q_{e}}{c}n_{\uparrow}[\textbf{v}_{\uparrow},\textbf{B}],\end{equation}
and
$$mn_{\downarrow}(\partial_{t}+\textbf{v}_{\downarrow}\nabla)\textbf{v}_{\downarrow}+\nabla p_{\downarrow}$$
\begin{equation}\label{SEAS EX Euler eq electrons spin DOWN} =q_{e}n_{\downarrow}\textbf{E}+\chi q_{e}^{2}\sqrt[3]{n_{\downarrow}}\nabla n_{\downarrow}+\frac{q_{e}}{c}n_{\downarrow}[\textbf{v}_{\downarrow},\textbf{B}],\end{equation}
where we present the explicit form of the Coulomb exchange interaction. The magnetic field $\textbf{B}$ in equations (\ref{SEAS EX Euler eq electrons spin UP}) and (\ref{SEAS EX Euler eq electrons spin DOWN}) is the external magnetic field.

The electric field in simplified Euler equations (\ref{SEAS EX Euler eq electrons spin UP}) and (\ref{SEAS EX Euler eq electrons spin DOWN}) is the quasi-static electric field. Hence it obeys the Poisson equation
\begin{equation}\label{SEAS EX div E} \nabla \textbf{E}=4\pi(en_{i0}-en_{e\uparrow}-en_{e\downarrow}),\end{equation}
and the eddy-free condition
\begin{equation}\label{SEAS EX rot E} \nabla\times \textbf{E}=0.\end{equation}

In equilibrium state we have $n_{i0}=n_{0e}=n_{0\uparrow}+n_{0\downarrow}$.

Considering quantum spin-1/2 plasmas researchers usually apply equation of state for unpolarized electrons
\begin{equation}\label{SEAS EX EqState unPol} p_{unpol}=\frac{(3\pi^{2})^{\frac{2}{3}}}{5}\frac{\hbar^{2}}{m}n^{\frac{5}{3}}, \end{equation}
see Ref. \cite{Landau v5 eq st}.

One can include the contribution of the spin polarization of the degenerate electron gas in the pressure in the single fluid model of the three dimensional electron gas \cite{MaksimovTMP 2001 b}, \cite{Andreev AoP 14}
\begin{equation}\label{SEAS EX eq of state Single fluid} p_{e}=\vartheta_{3D}(3\pi^{2})^{\frac{2}{3}}\frac{\hbar^{2}n_{e}^{\frac{5}{3}}}{5m_{e}},\end{equation}
where
\begin{equation}\label{SEAS EX vartheta} \vartheta_{3D}=\frac{1}{2}[(1+\eta)^{5/3}+(1-\eta)^{5/3}]\end{equation}
is the coefficient describing the spin polarization of the electron gas.

However, we deal with the separate evolution of the spin-up electrons and the spin-down electrons. Therefore, we apply the partial pressure caused by each species of electrons \cite{Andreev PRE 15}. Hence, we have
\begin{equation}\label{SEAS EX eq of state spin UP} p_{e\uparrow}=(6\pi^{2})^{\frac{2}{3}}\frac{\hbar^{2}n_{e\uparrow}^{\frac{5}{3}}}{5m_{e}},\end{equation} for the spin-up electrons, and
\begin{equation}\label{SEAS EX eq of state spin DOWN} p_{e\downarrow}=(6\pi^{2})^{\frac{2}{3}}\frac{\hbar^{2}n_{e\downarrow}^{\frac{5}{3}}}{5m_{e}},\end{equation}
for the spin-down electrons. Sum of $p_{e\uparrow}$ and $p_{e\downarrow}$ gives us $p_{e}$ presented by formula (\ref{SEAS EX eq of state Single fluid}) if we include the relation between concentrations $n_{\uparrow}=(1-\eta)n_{e}/2$ and $n_{\downarrow}=(1+\eta)n_{e}/2$.

Difference of the spin-up electron concentration and the spin-down electron concentration $\Delta n=n_{0\uparrow}-n_{0\downarrow}$ is caused  by the external magnetic field. Since electrons are negatively charged their spins have preferable direction opposite to the external magnetic field $\frac{\Delta n}{n_{0}}=\tanh(\frac{\mu_{e}B_{0}}{T_{Fe}})=-\tanh(\frac{\mid\mu_{e}\mid B_{0}}{T_{Fe}})$, $\eta=\mid\Delta n\mid/n_{0e}$, where $T_{Fe}=(3\pi^{2}n_{0e})^{2/3}\hbar^{2}/2m$ is the Fermi temperature in units of energy, so we do not write the Boltzmann constant.

\section{Perturbation evolution}

We consider the propagation of the non-linear perturbations parallel to the external magnetic field $\textbf{B}=B_{0}\textbf{e}_{z}$. Here, the plane wave soliton propagates parallel to the external magnetic field. Hence it parameters depend on the single coordinate and we have one dimensional perturbation. We focus our attention on the non-linear waves related to the SEAWs. We consider ions as the motionless positively charged background. To find the soliton solution we apply the perturbation technic developed by Washimi and Taniuti in Ref. \cite{Washimi PRL 66}. This technic is widely applied in recent research of wave phenomena in plasma physics. In this paper we apply it to find the spin-electron acoustic soliton.

In equations (\ref{SEAS EX cont eq electrons spin UP})-(\ref{SEAS EX Euler eq electrons spin DOWN}) we make transition to variables $\xi$ and $\tau$ defined as follows
\begin{equation}\label{SEAS EX def of xi} \xi=\varepsilon^{\frac{1}{2}}(z-Vt),\end{equation}
and
\begin{equation}\label{SEAS EX} \tau=\varepsilon^{\frac{3}{2}}t,\end{equation}
where $\varepsilon\ll1$ is a dimensionless parameter.

Following Ref. \cite{Washimi PRL 66} we introduce an expansion of the hydrodynamic parameters on small parameter $\varepsilon$
\begin{equation}\label{SEAS EX expansion of conc} n_{s}=n_{0s}+\varepsilon n_{1s}+\varepsilon^{2} n_{2s},\end{equation}
\begin{equation}\label{SEAS EX expansion of veloc} v_{sz}=0+\varepsilon v_{1sz}+\varepsilon^{2} v_{2sz},\end{equation}
and
\begin{equation}\label{SEAS EX expansion of phi} \phi=0+\varepsilon \phi_{1}+\varepsilon^{2} \phi_{2},\end{equation}
where $\phi$ is the potential of the electric field $\textbf{E}=-\nabla\phi$.

We substitute formulae (\ref{SEAS EX expansion of conc})-(\ref{SEAS EX expansion of phi}) in equations (\ref{SEAS EX cont eq electrons spin UP})-(\ref{SEAS EX div E}) In the leading order on the small parameter $\varepsilon$ we find the following relations between the perturbations of the particle concentrations, the velocity fields, and the potential of the electric field
\begin{equation}\label{SEAS EX relation between v1 and n1} v_{1s}=\frac{V}{n_{0s}}n_{1s},\end{equation}
\begin{equation}\label{SEAS EX n u 1 via phi 1} n_{1\uparrow}=  \frac{-en_{0\uparrow}\phi_{1}}{m_{e}(V^{2}-U_{\uparrow}^{2})},   \end{equation}
and
\begin{equation}\label{SEAS EX n d 1 via phi 1} n_{1\downarrow}=\frac{-en_{0\downarrow}\phi_{1}}{m_{e}(V^{2}-U_{\downarrow}^{2})}, \end{equation}
where we have applied the following notations
\begin{equation}\label{SEAS EX U up DEF} U_{\uparrow}^{2}=\frac{\hbar^{2}}{3m_{e}^{2}}(6\pi^{2}n_{0\uparrow})^{\frac{2}{3}},\end{equation}
and
\begin{equation}\label{SEAS EX U down DEF} U_{\downarrow}^{2}=\frac{\hbar^{2}}{3m_{e}^{2}}(6\pi^{2}n_{0\downarrow})^{\frac{2}{3}}-\frac{\chi e^{2}}{m_{e}}n_{0\downarrow}^{\frac{1}{3}}.\end{equation}

Below, we find that $V^{2}-U_{\uparrow}^{2}$ and $V^{2}-U_{\downarrow}^{2}$ have different signs. Hence perturbations $n_{1\uparrow}$ and $n_{1\downarrow}$ have different signs either.

The Poisson equation in the leading order on $\varepsilon$ appears as
\begin{equation}\label{SEAS EX Poisson eq 1} n_{1\uparrow}+n_{1\downarrow}=0. \end{equation}

Substituting formulae (\ref{SEAS EX n u 1 via phi 1}) and (\ref{SEAS EX n d 1 via phi 1}) in equation (\ref{SEAS EX Poisson eq 1}) we find the velocity of the perturbation propagation introduced in formula (\ref{SEAS EX def of xi}):
\begin{equation}\label{SEAS EX V full} V^{2}=\frac{1}{n_{0e}}(n_{0\uparrow}U_{\downarrow}^{2}+n_{0\downarrow}U_{\uparrow}^{2}). \end{equation}

We need to find the explicit form of perturbations $n_{1\uparrow}$, $n_{1\downarrow}$, and $\phi_{1}$. To this end, we consider the hydrodynamic equations (\ref{SEAS EX cont eq electrons spin UP})-(\ref{SEAS EX div E}) in the next order on small parameter $\varepsilon$. In this regime we find $n_{2s}$, $v_{2sz}$ in terms of $\phi_{1}$. We substitute expressions for the second order perturbations of the particle concentration $n_{2s}$ in the Poisson equation
\begin{equation}\label{SEAS EX} \partial_{\xi}^{2}\phi_{1}=4\pi e (n_{2\uparrow}+n_{2\downarrow})\end{equation}
obtained from (\ref{SEAS EX div E}) in the second order on $\varepsilon$.

\begin{figure}
\includegraphics[width=8cm,angle=0]{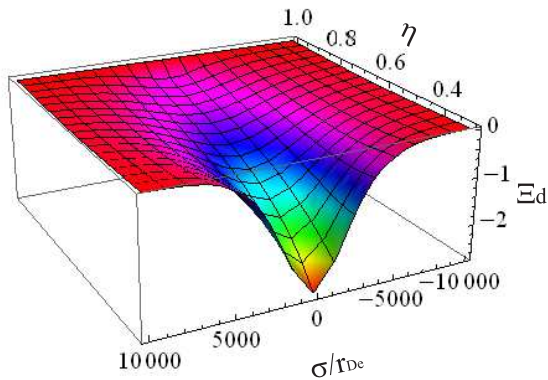}
\caption{\label{SEAS EX_Fig_Sol_d_noExch} (Color online) The
figure presents the spin polarization dependence of the soliton
profile of the spin-down electron concentration. We present the
soliton profile in terms of
$\Xi_{d}=(n_{1\downarrow}/n_{0e})/(\sqrt{3}U_{0}/v_{Fe})$, where
$n_{1\downarrow}$ appears at the substitution of solution
(\ref{SEAS EX solution for phi 1}) in formula (\ref{SEAS EX n d 1
via phi 1}).}
\end{figure}
\begin{figure}
\includegraphics[width=8cm,angle=0]{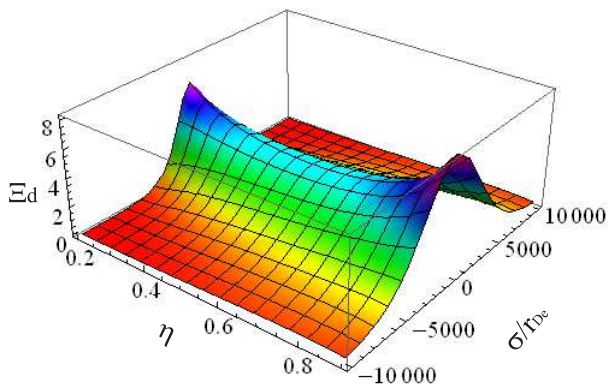}
\caption{\label{SEAS EX_Fig_Sol_u_noExch} (Color online) The
figure presents the spin polarization dependence of the soliton
profile of the spin-up electron concentration. Similarly to the
previous figure, we present the soliton profile in terms of
$\Xi_{u}=(n_{1\uparrow}/n_{0e})/(\sqrt{3}U_{0}/v_{Fe})$, where
$n_{1\uparrow}$ appears at the substitution of solution (\ref{SEAS
EX solution for phi 1}) in formula (\ref{SEAS EX n u 1 via phi
1}).}
\end{figure}
\begin{figure}
\includegraphics[width=8cm,angle=0]{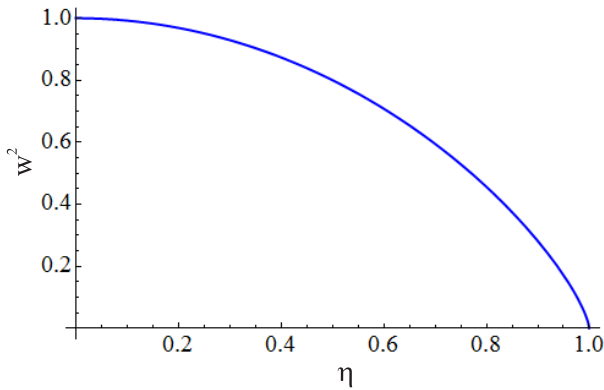}
\caption{\label{SEAS EX_Fig_vel_sq_noExch} (Color online) The figure shows the spin dependence of the square of the dimensionless perturbation velocity $w^{2}$ in self-consistent field approximation defined by formula (\ref{SEAS EX defin of w}).}
\end{figure}

Finally we find equation for the first order perturbation of the electric field potential
\begin{equation}\label{SEAS EX KdV} A\partial_{\tau}\phi_{1}+\partial_{\xi}^{3}\phi_{1}-B\partial_{\xi}\phi_{1}^{2}=0.\end{equation}
It is the KdV equation. Coefficients $A$ and $B$ are rather huge. We present and discuss them in Secs. IV and V. Let us mention that coefficient $A$ is positive $A>0$.

Applying the substitution $\varsigma=\xi-U_{0}\tau$ we find a soliton solution of KdV equation (\ref{SEAS EX KdV})
\begin{equation}\label{SEAS EX solution for phi 1} \phi_{1}=-\frac{3AU_{0}}{2B}\frac{1}{\cosh^{2}(\frac{1}{2}\sqrt{AU_{0}}\varsigma)}. \end{equation}
Parameter $U_{0}$ is the velocity of the soliton propagation. From formula (\ref{SEAS EX solution for phi 1}) we see that the amplitude of the soliton is proportional to the soliton velocity $U_{0}$, while the soliton width $\Delta$ is proportional to the inverse square root of the soliton velocity $U_{0}$.

We do not consider the ion contribution assuming ions as motionless. Therefore we have a limit on the perturbation velocity $V$.It should be larger than the sound velocity $v_{s}$: $V\gg v_{s}=\sqrt{\frac{m_{e}}{m_{i}}}v_{Fe}$, where
$v_{Fe}=(3\pi^{2}n_{0})^{1/3}\hbar/m$ is the Fermi velocity. Fig. \ref{SEAS EX_Fig_vel_sq_noExch} shows that at large spin polarization $\eta\rightarrow1$ the perturbation velocity becomes rather small $v_{Fe}\gg V$. In this regime the ion motion is essential. Therefore, we do not consider $\eta>0.9$ at the soliton study.

To find the spin electron acoustic soliton we apply a small perturbation evolution method. Consequently we have condition for the soliton amplitudes $n_{1s}\ll n_{0s}$. this condition gives restrictions for the velocity of the soliton propagation. For instance, considering the spin-up electrons we obtain
$n_{1\uparrow}\ll n_{0\uparrow}$  $\Rightarrow$ $U_{0}\ll \frac{m_{e}|B|(V^{2}-U_{\uparrow}^{2})}{A e \gamma_{\uparrow}}$.

\section{Spin electron acoustic soliton in the self-consistent field approximation}

In this section we present the analysis of the existence and properties  of the spin-electron acoustic soliton. To make this analysis simpler we consider the self-consistent field approximation. That means we drop the exchange interaction till the next section.

If we do not consider the exchange interaction we should drop the second term in the definition of $U_{\downarrow}^{2}$ (see formula (\ref{SEAS EX U down DEF})).

It is useful to represent the velocity of perturbation $V$, given by formula (\ref{SEAS EX V full}), in terms of the Fermi velocity. Hence, in the self-consistent field approximation, we find
$V^{2}\equiv\frac{1}{3}v_{Fe}^{2}\cdot w^{2}$
where
\begin{equation}\label{SEAS EX defin of w} w^{2}=\frac{1}{2}(1-\eta)^{\frac{2}{3}}(1+\eta)^{\frac{2}{3}}[(1-\eta)^{\frac{1}{3}}+(1+\eta)^{\frac{1}{3}}].\end{equation}
Coefficient $w^{2}$ describes the dependence of the perturbation velocity on the spin state of the electron gas.

Coefficients $A$ and $B$ in the KdV equation (\ref{SEAS EX KdV}), in the self-consistent field approximation, appear as follows
$$A= \frac{\sqrt{3}}{v_{Fe}r_{De}^{2}}\Biggl\{ \frac{1-\eta}{[w^{2}-(1-\eta)^{\frac{2}{3}}]^{2}} + \frac{1+\eta}{[w^{2}-(1+\eta)^{\frac{2}{3}}]^{2}} \Biggr\}$$
$$=\frac{\sqrt{3}}{v_{Fe}r_{De}^{2}} \Biggl\{ \frac{1}{(1-\eta)^{\frac{1}{3}}[\frac{1}{2}(1+\eta)^{\frac{2}{3}}[(1-\eta)^{\frac{1}{3}}+(1+\eta)^{\frac{1}{3}}]-1]^{2}}$$
\begin{equation}\label{SEAS EX}  +\frac{1}{(1+\eta)^{\frac{1}{3}}[\frac{1}{2}(1-\eta)^{\frac{2}{3}}[(1-\eta)^{\frac{1}{3}}+(1+\eta)^{\frac{1}{3}}]-1]^{2}} \Biggr\},\end{equation}
and
$$B=\frac{e}{m_{e}}\frac{3}{v_{Fe}^{2}r_{De}^{2}}\Biggl\{ \frac{\frac{1}{2}(1-\eta)}{[w^{2}-(1-\eta)^{\frac{2}{3}}]^{2}}\Biggl[\frac{1}{2}+\frac{w^{2}+\frac{1}{3}(1-\eta)^{\frac{2}{3}}}{w^{2}-(1-\eta)^{\frac{2}{3}}}\Biggr]$$
\begin{equation}\label{SEAS EX} + \frac{\frac{1}{2}(1+\eta)}{[w^{2}-(1+\eta)^{\frac{2}{3}}]^{2}}\Biggl[\frac{1}{2}+\frac{w^{2}+\frac{1}{3}(1+\eta)^{\frac{2}{3}}}{w^{2}-(1+\eta)^{\frac{2}{3}}}\Biggr] \Biggr\},\end{equation}
where $r_{De}=v_{Fe}/\sqrt{3}\omega_{Le}\sim n_{0e}^{-1/6}$ is the Debay radius. Coefficients $A$ and $B$ can be written as $A=A_{0}\cdot \tilde{A}(\eta)$ and $B=B_{0}\cdot \tilde{B}(\eta)$, where $A_{0}=\sqrt{3}v_{Fe}^{-1}r_{De}^{-2}$ and $B_{0}=3v_{Fe}^{-2}r_{De}^{-2}e/m_{e}$.

We see that coefficient $A$ does not depend on the equilibrium particle concentration.

The spin-electron acoustic soliton exists at the intermediate spin polarizations. It disappears at $\eta\rightarrow0$ and $\eta\rightarrow1$. Formally it corresponds to $A\rightarrow\infty$ at $\eta\rightarrow0$ and $\eta\rightarrow1$. Coefficient $B$ becomes infinite $B\rightarrow\infty$ at $\eta\rightarrow0$ and $\eta\rightarrow1$ either.

At the numerical analysis we consider the spin-electron acoustic soliton as the perturbations of the particle concentrations of the spin-up electrons and the spin-down electrons substituting solution (\ref{SEAS EX solution for phi 1}) in formulae (\ref{SEAS EX n u 1 via phi 1}) and (\ref{SEAS EX n d 1 via phi 1}).

The spin-electron acoustic soliton, for instance, for the spin-down electrons, arises as $n_{1d}=n_{0d}(1.5\tilde{A}/\tilde{B})(\sqrt{3}U_{0}/v_{Fe})[w^{2}-(1+\eta)^{2/3}]^{-1}\cosh^{-2}(\sigma/2\Delta)$, where $\Delta=\sqrt{v_{Fe}/\sqrt{3}U_{0}\tilde{A}}r_{De}$ is the soliton width.

Relative perturbations of the particle concentrations $n_{1s}/n_{0s}$ are proportional to the velocity of the soliton propagation $U_{0}$ in units of the Fermi velocity $v_{Fe}$. Presenting soliton profiles in figures we apply the effective amplitude $\Xi_{s}=(n_{1s}/n_{0e})/(\sqrt{3}U_{0}/v_{Fe})$.

Fig. (\ref{SEAS EX_Fig_Sol_d_noExch}) (Fig. (\ref{SEAS EX_Fig_Sol_u_noExch})) shows that the particle concentration perturbations are negative (positive) for the spin-down (spin-up) electrons. Applying notations accepted in some areas of the condensed matter physics and optics we can call the negative (positive) solitonic perturbations as the dark (bright) soliton.

For the particle concentration $n_{0e}=10^{23}$ cm$^{-3}$ we have $v_{Fe}/\sqrt{3}=7.9\times10^{7}$ cm/s. Choosing $U_{0}=10^{-8}v_{Fe}$ we obtain $n_{1s}=0.5(1\pm\eta)10^{15}\Xi_{s}$, where $\Xi_{s}$ is shown in Fig. (\ref{SEAS EX_Fig_Sol_d_noExch}) for the spin-down electrons and in Fig. (\ref{SEAS EX_Fig_Sol_u_noExch}) for the spin-up electrons.
Factor $0.5(1\pm\eta)$ appears due to the application of $n_{0e}$ in the definition of $\Xi_{s}$.
Figs. (\ref{SEAS EX_Fig_Sol_d_noExch}) and (\ref{SEAS EX_Fig_Sol_u_noExch}) demonstrate that the soliton width $\Delta$ is of order of $10^4 r_{De}=5\times10^{-5}cm$. Increasing the soliton velocity $U_{0}$ we decrease the soliton width $\Delta$. Since the soliton width $\Delta$ should be larger then the average interparticle distance $\Delta\gg n_{0e}^{-1/3}$ we obtain that $U_{0}\ll 10^{-2} v_{Fe}$ and $n_{1s}\ll 10^{21}$ cm$^{-3}$.

Amplitude of the dark soliton in the subsystem of spin-down electrons increases with the decrease of the spin polarization (\ref{SEAS EX_Fig_Sol_d_noExch}). Amplitude of the bright soliton in the subsystem of spin-up electrons has nonmonotonic dependence on the spin polarization (\ref{SEAS EX_Fig_Sol_u_noExch}). It has the minimum at the intermediate spin polarization $\eta_{0}=0.54$. The amplitude increases at change of the spin polarization $\eta$ from $\eta_{0}$ towards smaller or larger values. At change of the spin polarization from $\eta_{0}=0.54$ to $\eta=0.1$ and $\eta=0.9$ the amplitude increases from 4 to 6.7 and 8.2 correspondingly.

We present analysis of the soliton characteristics in the self-consistent field approximation. In the next section we include the contribution of the Coulomb exchange interaction in the spin-electron acoustic soliton propagation.

\section{Contribution of exchange interaction in spin electron acoustic soliton}

The spin dependence of the Coulomb exchange interaction force field in the electron gas is calculated in Ref. \cite{Andreev AoP 14}. Significant role of the Coulomb exchange interaction at rather large spin polarization is demonstrated there.

To underline the fact that, in this section, we consider the Coulomb interaction beyond the self-consistent field approximation we present the explicit form of the perturbation velocity obtained in Sec. III (see formula (\ref{SEAS EX V full}))
$ V^{2}=n_{0\uparrow}U_{\downarrow}^{2}+n_{0\downarrow}U_{\uparrow}^{2}/n_{0e}, $
where $n_{0e}=n_{0\uparrow}+n_{0\downarrow}$, functions $U_{\uparrow}^{2}$ and $U_{\downarrow}^{2}$ are defined by formulae (\ref{SEAS EX U up DEF}) and (\ref{SEAS EX U down DEF}), correspondingly.

Coefficient $A$ of the KdV equation (\ref{SEAS EX KdV}) at the account of the Coulomb exchange interaction appears as
\begin{equation}\label{SEAS EX} A=2V \Biggl\{ \frac{\omega_{L\uparrow}^{2}}{(V^{2}-U_{\uparrow}^{2})^{2}} +\frac{\omega_{L\downarrow}^{2}}{(V^{2}-U_{\downarrow}^{2})^{2}} \Biggr\}.\end{equation}
The exchange interaction between spin-down electrons being in states occupied by single electron modifies $U_{\downarrow}^{2}$. The account of the Coulomb exchange interaction does not change sign of $A$, so we have $A>0$.

The contribution of the exchange interaction in coefficient
$$B=\frac{e}{m_{e}} \Biggl\{ \frac{\omega_{L\uparrow}^{2}}{(V^{2}-U_{\uparrow}^{2})^{2}} \Biggl[\frac{1}{2}+\frac{V^{2}+\frac{1}{3}U_{\uparrow}^{2}}{V^{2}-U_{\uparrow}^{2}}\Biggr]$$
$$+\frac{\omega_{L\downarrow}^{2}}{(V^{2}-U_{\downarrow}^{2})^{2}} \Biggl[\frac{1}{2}+\frac{V^{2}+\frac{1}{3}U_{\downarrow}^{2}}{V^{2}-U_{\downarrow}^{2}}\Biggr]$$
\begin{equation}\label{SEAS EX B with Exchange}  +\frac{\omega_{L\downarrow}^{2}}{(V^{2}-U_{\downarrow}^{2})^{3}}\frac{1}{6}\frac{\chi e^{2} n_{0d}^{\frac{1}{3}}}{m_{e}} \Biggr\}\end{equation}
reveals in several ways. We find modifications of $U_{\downarrow}^{2}$ (see the second term in formula (\ref{SEAS EX U down DEF})) and $V^{2}$ (compare formulae (\ref{SEAS EX V full})). Moreover, we find an extra term (the last term in formula (\ref{SEAS EX B with Exchange})) in coefficient $B$.

\begin{figure}
\includegraphics[width=8cm,angle=0]{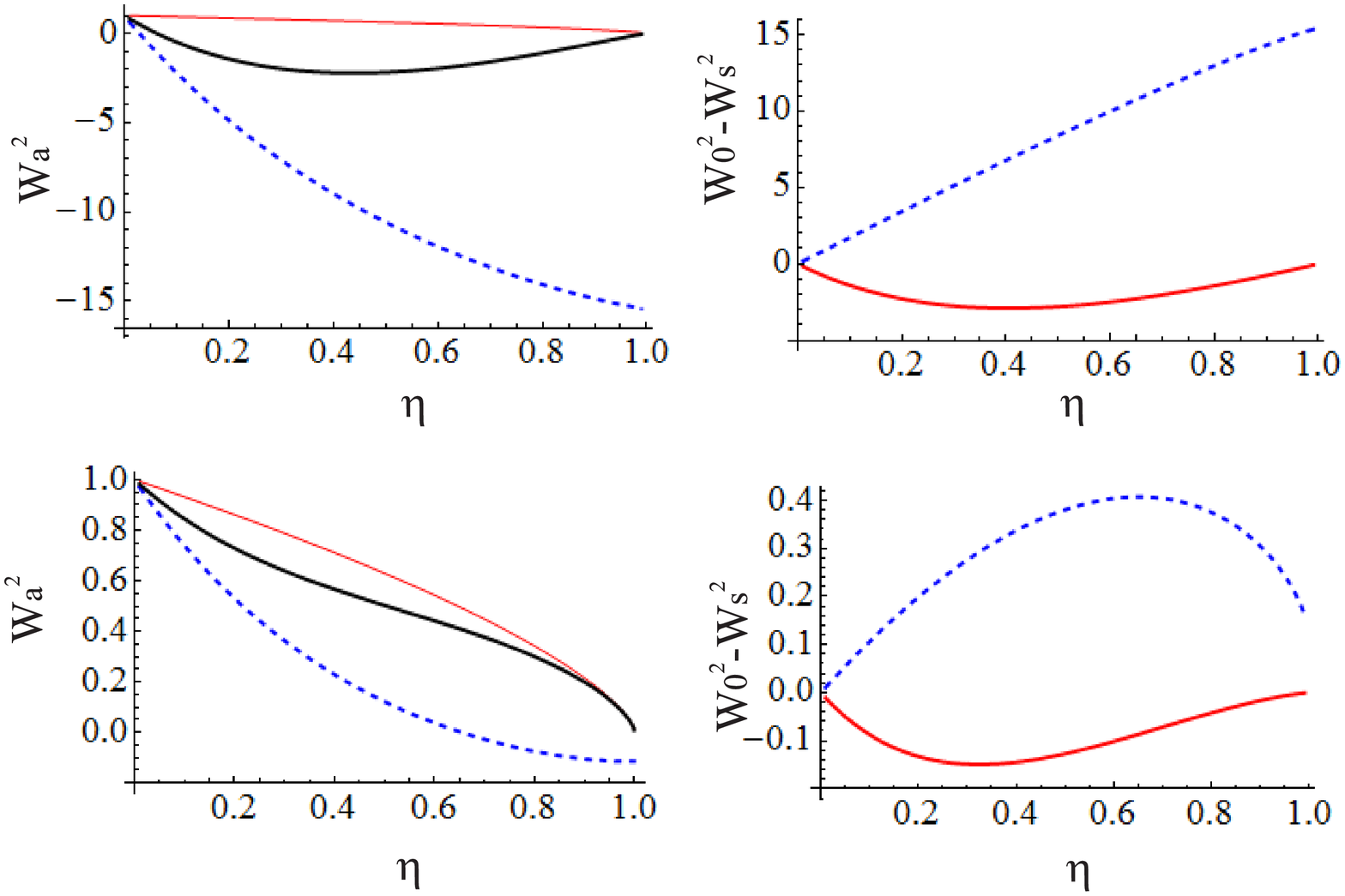}
\includegraphics[width=8cm,angle=0]{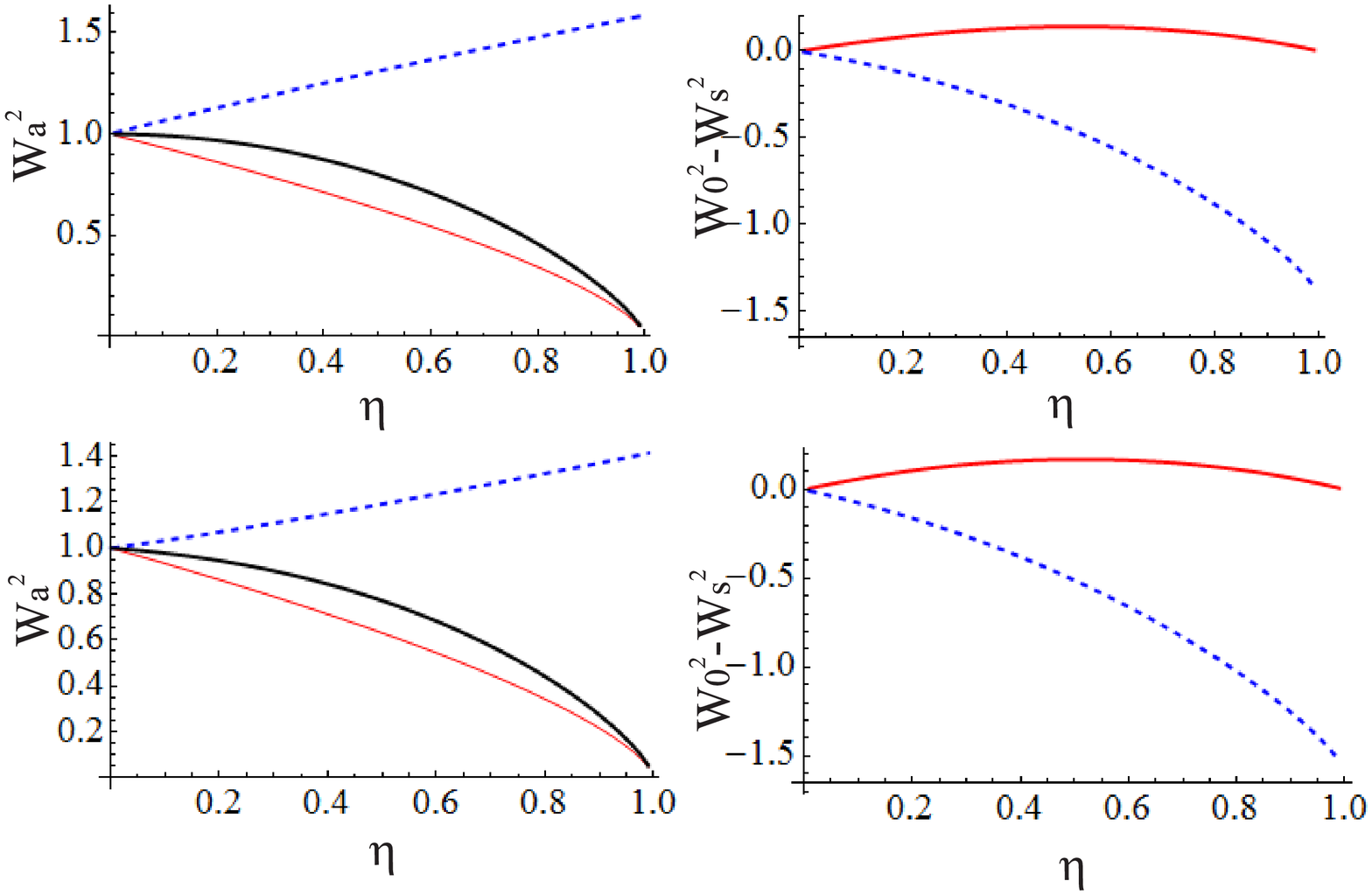}
\caption{\label{SEAS EX_velocities-diff_regimes} (Color online) The left column on the figure shows the dependencies of $W_{0}^{2}=3V^{2}(\eta)/v_{Fe}^{2}$ (thick black solid line), $W_{1}^{2}=3U_{\uparrow}^{2}(\eta)/v_{Fe}^{2}$ (thin red solid line), $W_{2}^{2}=3U_{\downarrow}^{2}(\eta)/v_{Fe}^{2}$ (blue dashed line) at different parameters of the electron gas (in order from the upper picture to the lower picture we have $n_{0}=10^{21}$ cm$^{-3}$ with the exchange interaction, $n_{0}=10^{24}$ cm$^{-3}$ with the exchange interaction, $n_{0}=10^{27}$ cm$^{-3}$ with the exchange interaction, any $n_{0}$ with no account of the exchange interaction). In this figure $W_{a}^{2}$ stands for $\{W_{0}^{2}, W_{1}^{2}, W_{2}^{2}\}$. The right column on the figure shows differences of the velocity squares presented in denominators in formulae (\ref{SEAS EX n u 1 via phi 1}) and (\ref{SEAS EX n d 1 via phi 1}) $W_{0}^{2}-W_{1}^{2}$ (red solid line) and $W_{0}^{2}-W_{2}^{2}$ (blue dashed line). In the figure we apply notation $W_{s}^{2}$ for $W_{1}^{2}$ and $W_{2}^{2}$. Regimes for differences $W_{0}^{2}-W_{s}^{2}$ in the right column correspond to the regimes for $W_{a}^{2}$ in the left column.}
\end{figure}
\begin{figure}
\includegraphics[width=8cm,angle=0]{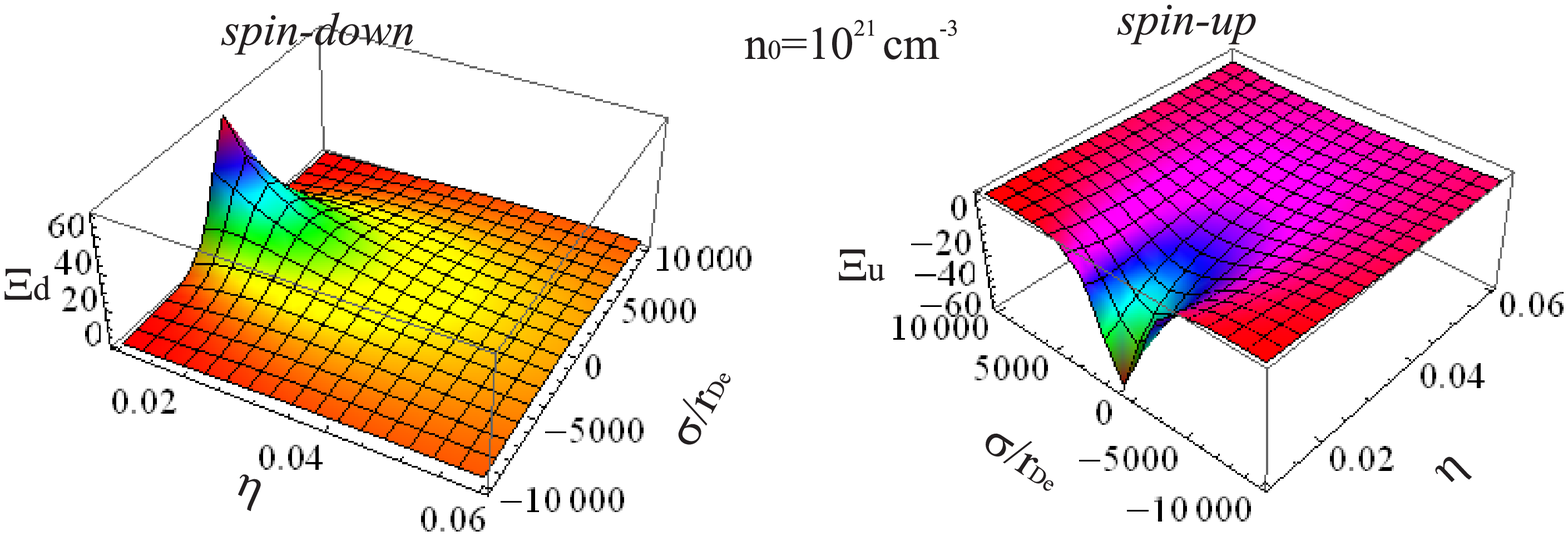}
\includegraphics[width=8cm,angle=0]{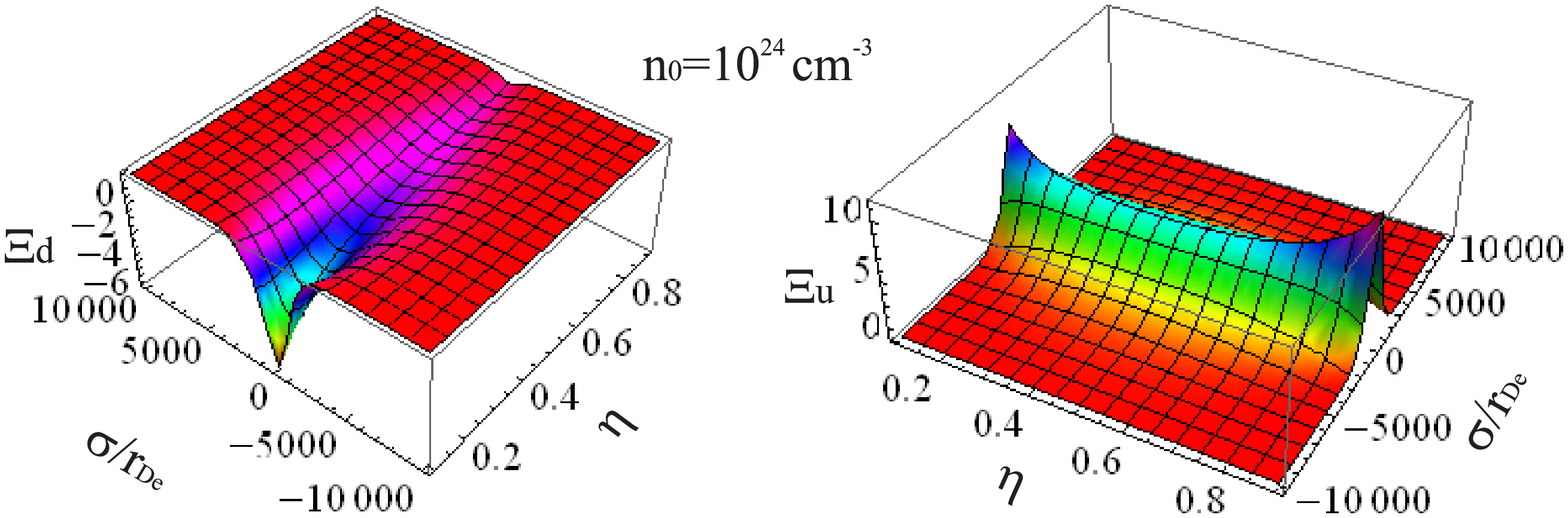}
\includegraphics[width=8cm,angle=0]{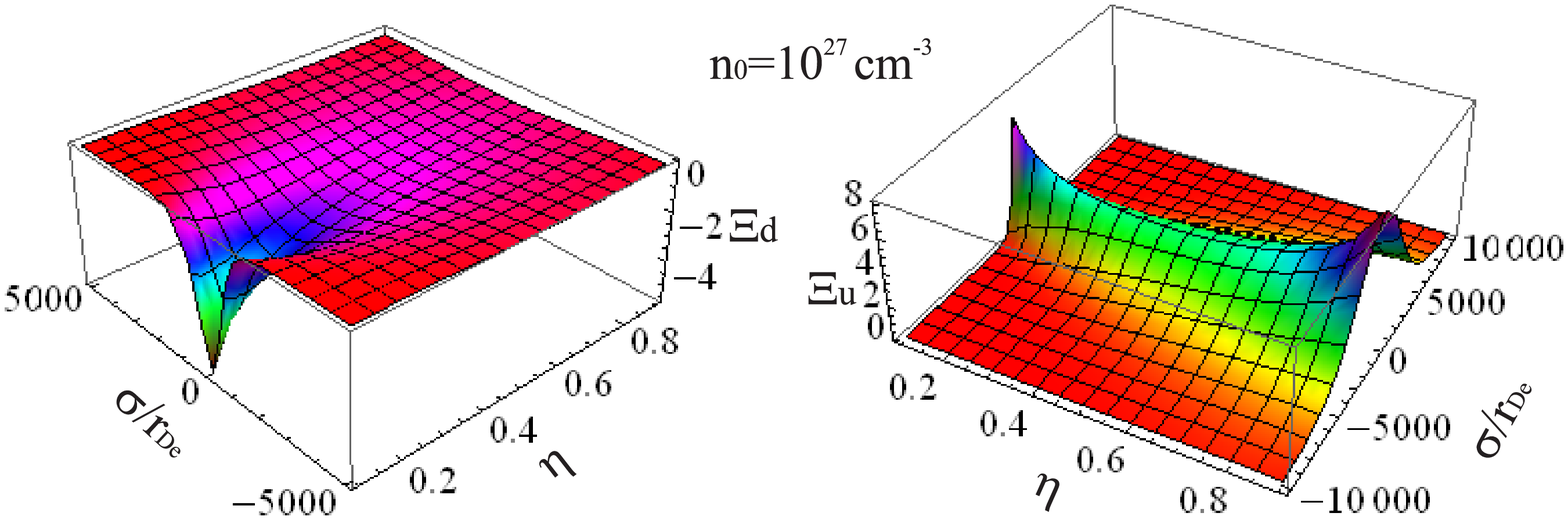}
\caption{\label{SEAS EX_solitons with Exch} (Color online) The figure shows the soliton profiles for spin-down (the left column) and spin-up (the right column) electron concentrations for different equilibrium concentrations of electrons presented in the figure.   We present the soliton profile in terms of $\Xi_{s}=(n_{1s}/n_{0s})/(\sqrt{3}U_{0}/v_{Fe})$, where $n_{1s}$ appears at the substitution of solution (\ref{SEAS EX solution for phi 1}) in formulae (\ref{SEAS EX n d 1 via phi 1}) and (\ref{SEAS EX n u 1 via phi 1}).}
\end{figure}

Coefficient $B$ is positive in the self-consistent field approximation $B_{SCF}>0$. The dependence of the $B_{SCF}$ on the electron concentration is located in $B_{0}$. If we include the exchange interaction the behavior of the coefficient $B$ becomes rather complicate. We find a dependence of $B/B_{0}$ on the equilibrium electron concentration $n_{0e}$.

This dependence reveals in properties of the soliton profile described below. We can track this dependence at an intermediate spin polarization $\eta'=0.5$. We see that coefficient $B$ is positive $B>0$ in the regime of rather large concentrations $n_{0e}\sim10^{27}$ cm$^{-3}$, which is similar to the self-consistent field approximation. At $\tilde{n}_{0}\approx5.4\times10^{24}$ cm$^{-3}$ we find fast increase of $B$ ($B\rightarrow +\infty$). After point $\tilde{n}_{0}$ the coefficient $B$ increases from minus infinity up to the zero value $B=0$ at $n_{0}'=0.94\times 10^{23}$ cm$^{-3}$. At smaller concentrations the coefficient $B$ is positive. We see that the coefficient $B$ as a function of the electron concentration demonstrates the hyperbolic dependence. Since $B$ is in the denominator of the soliton amplitude, the soliton amplitude tents to zero at $\tilde{n}_{0}$. Hence, the soliton does not exist near this concentration. At $n_{0e}=n_{0}'$ coefficient $B$ vanishes. Consequently, the soliton amplitude approach infinity. Therefore, the perturbation method, we apply to find the spin-electron acoustic soliton, cannot give any information about soliton behavior near this point.

For the formation of the soliton we need to have waves with the stable linear spectrum. Hence, we need to have a positive square of the perturbation velocity given by formula (\ref{SEAS EX V full}), which is a combination of $U_{\uparrow}^{2}$ and $U_{\downarrow}^{2}$. In the self-consistent field approximation considered in the previous section the spectrum is stable for all values of parameters. If we include the exchange interaction situation changes. Parameter $U_{\downarrow}^{2}$ contains a negative term caused by the exchange interaction (see formula (\ref{SEAS EX U down DEF})). Therefore, $V^{2}$ can become negative. To find areas of positive $V^{2}$ we present Fig. \ref{SEAS EX_velocities-diff_regimes}. In the left-hand column we depict $V^{2}$, $U_{\uparrow}^{2}$ and $U_{\downarrow}^{2}$ as the functions of the spin polarization $\eta$ at the different equilibrium electron concentrations $n_{0e}$. The lower row of pictures is obtained in the self-consistent field approximation to provide the comparison with the results of the previous section.

The results of the previous section show that sign of quantities $V^{2}-U_{\uparrow}^{2}$ and $V^{2}-U_{\downarrow}^{2}$ define profiles of the spin-electron acoustic solitons of the concentration of the spin-up and spin-down electrons. Therefore, we present these quantities in the right-hand column in Fig. \ref{SEAS EX_velocities-diff_regimes}.

Fig. \ref{SEAS EX_velocities-diff_regimes} shows that at $n_{0e}=10^{27}$ cm$^{-3}$ results including the exchange interaction have small, but noticeable, difference with the results of the self-consistent field approximation. Hence, in this regime, we find the soliton profiles (see the lower row in Fig. \ref{SEAS EX_solitons with Exch}) similar to the profiles found in the previous section (see Figs. \ref{SEAS EX_Fig_Sol_d_noExch} and \ref{SEAS EX_Fig_Sol_u_noExch}). However, we see the contribution of spin-down electrons presented by blue dashed lines in Fig. \ref{SEAS EX_velocities-diff_regimes} on 14 percent, approximately, due to the account of the exchange interaction. It reveals in the amplitude of the spin-down soliton.
Linear spectrum is stable in this regime $V^{2}>0$, and the parameters $V^{2}-U_{\uparrow}^{2}$ and $V^{2}-U_{\downarrow}^{2}$ have positive and negative signs correspondingly ($U_{\uparrow}^{2}<V^{2}<U_{\downarrow}^{2}$). In this regime the coefficient $B$ is positive $B>0$. Consequently, we find the dark soliton in the spin-down electron concentration and the bright soliton in the spin-up electron concentration.

Decreasing the equilibrium concentration of electrons down to $n_{0e}=10^{24}$ cm$^{-3}$ we find that the spectrum is stable ($V^{2}>0$) for all $\eta\in [0,1]$. However, relative values of $V^{2}$, $U_{\uparrow}^{2}$, $U_{\downarrow}^{2}$ are changed. Velocity $U_{\downarrow}$ becomes the smallest of them. So, we have $U_{\downarrow}^{2}<V^{2}<U_{\uparrow}^{2}$.
Consequently the signs of parameters $V^{2}-U_{\uparrow}^{2}$ and $V^{2}-U_{\downarrow}^{2}$ are changed either.
In this regime the coefficient $B$ is negative $B<0$. Therefore, we find the dark soliton in the spin-down electron concentration and the bright soliton in the spin-up electron concentration.

At $n_{0e}=10^{21}$ cm$^{-3}$ the linear spectrum becomes unstable in the wide range of the spin polarization. We find the stability interval at $\eta\in (0.01,0.06)$. In this regime the coefficient $B$ is positive $B>0$.
Consequently, in the area of stability, we find the bright soliton in the spin-down electron concentration and the dark soliton in the spin-up electron concentration (see the upper row in Fig. \ref{SEAS EX_solitons with Exch}).

\section{Conclusions}

Existence of the spin-electron acoustic soliton has been discovered. Its existence closely related to the spin electron acoustic waves recently obtained in Ref. \cite{Andreev PRE 15}. The SEAW has linear spectrum, so it resembles similarity to the ion-acoustic wave, but the SEAW has larger frequencies. The balance between the dispersion and nonlinearity in the SEAWs of small amplitude allows to form a soliton solution obtained in this paper and called the spin-electron acoustic soliton.

To study the non-linear spin-electron acoustic waves we have developed a generalization of the separate spin evolution quantum hydrodynamics. This generalization includes the Coulomb exchange interaction. The exchange interaction appears from the interaction of the spin-down electrons being in the states occupied by one electron only. This mechanism was considered in the single fluid model of  electrons. In this paper we have adopted it for the SSE-QHD.

We have considered the spin-electron acoustic soliton in two regimes. First of all we have considered it in, rather simple, regime of the self-consistent field approximation. We have found that the spin-electron acoustic soliton shows itself as the dark soliton of the spin-down electron concentration and the bright soliton of the spin-up electron concentration. The soliton shows similar behavior for all equilibrium concentrations of electrons.

The second regime of the spin-electron acoustic wave study includes the Coulomb exchange interaction.

The exchange interaction significantly change properties of the spin-electron acoustic soliton. Strong dependence of the soliton properties reveals in this regime. At equilibrium concentration $n_{0e}=10^{21}$ cm$^{-3}$ we have found, in opposite to the self-consistent field approximation, the bright soliton of the spin-down electron concentration and the dark soliton of the spin-up electron concentration, existing in the narrow interval at rather small spin polarizations $\eta\in(0.01,0.06)$.

The increase of the equilibrium electron concentration increases the area of the soliton existence. At $n_{0e}=10^{24}$ cm$^{-3}$ we have found the existance of the spin-electron acoustic soliton at $\eta\in(0.01,0.99)$. At $n_{0e}\geq10^{24}$ cm$^{-3}$ we have obtained
the dark soliton of the spin-down electron concentration and the bright soliton of the spin-up electron concentration. It is in the agreement with the self-consistent field approximation. However, the parameters of the soliton at the account of the exchange interaction at $n_{0e}\in [10^{24},10^{27}]$ cm$^{-3}$ differ from the results of the self-consistent field approximation.


\begin{acknowledgements}
The author thanks Professor L. S. Kuz'menkov for fruitful discussions. The author thanks the Dynasty foundation for financial support.
\end{acknowledgements}

\end{document}